\documentclass{article}
\usepackage{amsmath}
\usepackage{amsfonts}
\usepackage{latexsym}
\usepackage{amscd}
\usepackage{amssymb}
\usepackage{graphicx}
\usepackage{float}

\def \n {\noindent}

\restylefloat{float}

\newtheorem{theorem}{Theorem}[section]

\newtheorem{remark}{Remark}[section]

\title{Global stability for $SIR$ and $SIRS$ models with differential mortality}

\author{ P. Adda and D. Bichara \\
 \'Equipe-projet Masaie et Universit\'e de Metz\\
LMAM (UMR CNRS 7122)\\
I.S.G.M.P Bat A, Ile du Sauly, 57045 Metz Cedex 01,\\
France\\
\{philippe.adda, derdei.bichara\}@inria.fr\\
}
\date{}
\begin{document}
\maketitle
\abstract{We consider  $SIR$ and $SIRS$ models with differential mortality. Global stability of equilibria is established by using Lyapunov's method.}

\n
{\bf Keywords: } Nonlinear dynamical systems, global stability, Lyapunov methods.

\n 
{\bf AMS subject classification} : 34A34, 34D23, 34D40, 92D30

\n
\section{Introduction} 
\n
The SIR model is a classical model in mathematical epidemiology. Particularly  Kermack and McKendrick \cite{KMcK1927} use a SIR  model to prove the existence of threshold. The model of Kermack and McKendrick is without demography, i.e. without vital dynamics. The classic SIR models  are very important as conceptual models (similar to predator-prey and competing
species models in ecology). The SIR epidemic modeling yields the useful concept of the threshold.

When a vital dynamic is introduced the asymptotic behavior changes. When the death rates are equal in each compartments S,I and R, and equal to the birth rate, the global stability has been solved in \cite{MR1026035,0326.92017}. Actually, the total population is constant, hence the system reduces to a two dimensional system. Then using phase plane methods  (Poincar\'e-Birkhoff) and Lyapunov functions the global stability is obtained.

Models with a variable total population size are often more difficult to analyze
mathematically because the population size is an additional variable which is governed
by a differential equation

The global stability using Lyapunov functions of SIR model with a total constant population  is proved in \cite{1022.34044}. However in this model the death rates of $S$ and $I$ are equal and the death rate of the removed compartment is adjusted relatively to the death rate of S and the constant birth rate. This adjustment is just done to have a constant  total population.
This is a little bit artificial. The model with constant population simplifies in two important ways :

\begin{itemize}
\item The mass action law  $\displaystyle \frac{SI}{N}$ reduces to a bilinear law $\tilde \beta S \,I$
\item The system is a two-dimensional system.
\end{itemize}

In this paper we propose a more realistic model, with constant population. We suppose, which is more or less observable, that the natality compensates for the mortality. Our model can deal with different death rates, and particularly with a  over-mortality from the disease.

We denote $\mathcal R_0$ the basic reproduction number. It is  defined as the expected number of new cases of infection caused by a typical infected in a population susceptible \cite{Diekman2000, R0vdd}. We prove in this paper the global stability of disease free equilibrium (DFE) if $\mathcal R_0\leq1$ and  that there exists a unique endemic equilibrium (EE) if $\mathcal R_0>1$, which is globally asymptotically stable on the domain minus the stable manifold of the DFE.

The stability analysis of classical SIR model is well know since 1976 \cite{0326.92017, MR2002c:92034}. The reason was that study of stability for these models reduce to the study of 2-dimensionnal systems, hence phase methods can be used: Poincar\'e-Bendixon theorem. Periodic orbits are ruled out using Dulac criteria or a condition of Busenberg and Van Den Driessche \cite{BusVdd}.

In the recent litterature, the Lyapunov method is successfuly used to prove the global stability of endemic equilibrium. This method consists to find one function, called Lyapunov function and  usually denoted by $V$, positive definite and its derivative along trajectories is negative definite. If the derivative $\dot V$ is only negative , the LaSalle's invariance principle extend the Lyapunov method in particular cases. This Lyapunov function is very difficult to exhibit. However, the class of  Lyapunov function 
$$V=\sum_{i=1}^{n}a_i\left(x_i-\bar x_i\log x_i \right)$$ is used.  This function has a long history of application to Lotka-Voltera models and was discovered by Voltera himself, although he did not use the vocabulary and the theory of Lyapunov functions. In 2002, Korobeinikov and Wake use this type of function to prove the global stability for $SIR$, $SIRS$ and $SIS$ models \cite{1022.34044} and in 2004, for $SEIR$ and $SEIS$ model \cite{Korob04} and give a simply proof  of the result of  Li and Muldowney \cite{Li95}.

We give a brief outline of the paper. In section 2, we formulate the model and study the global stability of the  DFE if $\mathcal R_0\leq1$. In section 3, we study the global stability of endemic equilibrium if $\mathcal R_0>1$. In section 4,  we deduce results of the next section to SIRS model. In finally we conclude in section 5.

\section{Model formulation}
We consider a population $N$ divided into classes of susceptible, infectious and removed individuals, with numbers at time $t$ denoted by $S(t)$, $I(t)$ and $R(t)$ respectively, that is $N=S(t)+I(t)+R(t)$.   We assume that there is no vertical transmission, then all offsprings are susceptibles. We assume that the natality  $\Lambda$ compensates for the deaths. Then $\Lambda =\mu_1S+\mu_2I+\mu_3R$. The parameter $\gamma$ is the rate of recovery. Note that, in our model the disease confers a permanent immunity. The parameter $\beta$ is the effective per capita contact rate of infective individuals. We modelize the contact by the classical law of  mass action mass. We have the following flow  graph:
{
  \begin{figure}[ht]
  \begin{center}
{ \includegraphics[scale=0.8]{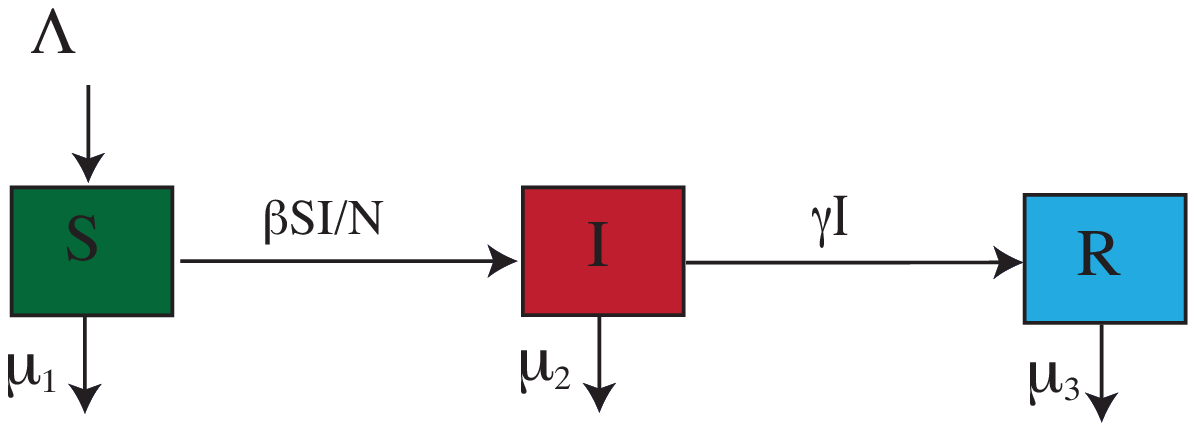}}
 \end{center}
 \end{figure}
 }
\\
\noindent The dynamic of this model is given by the following system:

\begin{equation} \label{0}
\left\{\begin{array}{lll}
\dot{S}=\Lambda-\beta\frac{S \,I}{N}-\mu_1 \,S\\
\dot{I}=\beta\frac{S \,I}{N}-\mu_2I - \gamma \,  I\\
\dot{R}=\gamma\, I - \mu_3 \, R 
\end{array}\right.
\end{equation}

\noindent Which reduces to 
\begin{equation} \label{1}
\left\{\begin{array}{lll}
\dot{S}=-\beta\frac{S\,I}{N}+\mu_2\,I+\mu_3\,R\\
\dot{I}=\beta\frac{S\,I}{N}-\mu_2\,I - \gamma\, I\\
\dot{R}=\gamma \,I - \mu_3\, R 
\end{array}\right.
\end{equation}
The population size is constant, so that $S+I+R=N$, then we can omit the equation of removed population. We obtain the two-dimensional system:
$$
\left\{\begin{array}{ll}
\dot{S}=-\beta\frac{SI}{N}+\mu_2I+\mu_3\left(N-S-I\right)\\
\dot{I}=\beta\frac{SI}{N}-\left(\mu_2 + \gamma \right)I
\end{array}\right.
$$

\noindent
For simplicity we can consider the prevalence, i.e. the proportions.

If we denote $\frac{S}{N}$, $\frac{I}{N}$, the susceptible and infectious fractions, again by $S$ and $I$. Then the system (\ref{1}) is reduced to

\begin{equation}\label{2}
\left\{\begin{array}{ll}
\dot{S}=\mu_3 + (\mu_2-\mu_3)I -\mu_3S-\beta SI\\
\dot{I}=\beta SI-\left(\mu_2 + \gamma \right)I
\end{array}\right.
\end{equation}

\medskip
\noindent
We have $0 \leq S $, $0 \leq I$ and $S+I \leq1$. The biological domain of this two-dimensional system is the standard simplex.

\medskip
\noindent
The set $\Omega=\left\{\left(S,I\right):S\geq0; I\geq0; S+I\leq1\right\}$ is a positively invariant compact set for (\ref{2}).
The system is well posed.

\medskip
\noindent
The basic reproduction ratio is given by

 $$\mathcal R_0=\frac{\beta}{\mu_2+\gamma}.$$

\subsection{Stability of DFE}

\noindent
System (\ref{2}) has a 
 disease free equilibrium state, which   is given by  $\left(S^\ast, 0\right)=\left(1,0\right).$ 
 
 \medskip
\begin{theorem}
If $\mathcal R_0\leq1$ then the DFE is globally asymptotically stable on $\Omega$.
\end{theorem}
\textbf{Proof:} \\
We consider the Lyapunov-LaSalle  function $V(S,I)=I$. We have:
\begin{eqnarray}
\dot V &=& \dot I \nonumber\\
       &=& \beta SI-\left(\mu_2 + \gamma \right)I \nonumber\\
       &=& I\left(\mathcal R_0S-1\right)\left(\mu_2+\gamma\right)\nonumber\\
       &\leq& 0 \nonumber
\end{eqnarray}
Furthermore $\dot V=0$ if $I=0$ or $S=S^\ast$ and $\mathcal R_0=1$. Hence the largest invariant set contained in the set $\displaystyle \mathcal L=\left\{\left(S,I\right)\in\Omega\;\;/\;\;\dot{V}\left(S,I\right)=0\right\}$ is reduced to the DFE.
Since we are in a compact positively invariant set,  by the LaSalle's Invariance Principle \cite{LaSLef61,Bhatia70} , the DFE is globaly asymtotically stable in $\Omega$.

\medskip
\begin{remark}

Unlike Lyapunov's theorems, 
LaSalle's principle does not require the function $V(x)$ to be 
positive definite . If  the largest invariant set $M$, 
contained in the set $E$ of points where $\dot{V}$ vanishes, is reduced to 
the equilibrium point, i.e. if  $M=\{x_{0}\}$, the LaSalle's principle 
allows to conlude that the equilibrium is attractive.
But a drawback of Lasalle's principle, when 
significant,  is that it 
proves only the attractivity of the equilibrium point.
It is well known that in the nonlinear case attractivity 
does not imply stability.
But when the function $V$ is not positive definite, Lyapunov stability 
must be proven.
This is why LaSalle's principle is often misquoted. 
Some additional condition enables, with LaSalle's principle,  to ascertain asymptotic stability. 
To obtain stability from LaSalle's principle some additional work is 
needed. The most complete results, in the direction of Lasalle's principle to 
prove asymptotic stability, have been obtained by LaSalle himself 
({\bf LaSalle}:\cite{Las68}, in 1968, completed in 1976 
\cite{LaSalleNLATMA76}  .)
\end{remark}

\section{Global Stability of endemic equilibrium}

\noindent 
An equilibrium for system (\ref{2}), different from the DFE,    is given by $\left(\bar S,\bar I\right)$, where

$$\bar S=\frac{\mu_2+\gamma}{\beta}= \frac{1}{\mathcal R_0}  \;\;\; {  \rm and   } \;\;\;  \bar I=\frac{\mu_3}{\mu_3+\gamma}\left(1-\frac{1}{\mathcal R_0}\right)$$

\noindent
This equilibrium is in the simplex, i.e., $0 \leq  \bar S$, $0\leq \bar I $ and $\bar S +\bar I \leq 1$ iff $\mathcal R_0 >1$.

\noindent
Clearly $0 \leq \bar I$ is equivalent to $\mathcal R_0 \geq 1$. Now we can write

$$\bar S +\bar I = \frac{\frac{\gamma}{\mathcal R_0}+\mu_3}{\gamma+\mu_3}$$ 

\noindent
When $\mathcal R_0=1$ this equilibrium coincides with the DFE. 
Then there is an unique equilibrium in the interior of the simplex iff $\mathcal R_0 >1$.

\medskip 
 \begin{theorem}
 If $\mathcal R_0>1$, the DFE is unstable  and there exists a unique endemic equilibrium $(\bar S, \bar I)$ and this endemic equilibrium is globally asymptotically stable on the domain $\Omega \setminus \ [0,1] \times \{0\}$. In other words on the simplex minus the stable manifold of the DFE
 \end{theorem}

\noindent \textbf{Proof:} \\
 
 \noindent
 When $\mathcal R_0 >1$ the instability of the DFE comes from \cite{Diekman2000}.
 
 Let $\Omega_1$ the set defined by  $\displaystyle \Omega_1=\left\{(S,I)/\;S\geq\frac{\mu_2-\mu_3}{\beta},\;I\geq0,\;S+I\leq1\right\}$.  The set $\Omega_1$ is a compact positively invariant. We Consider on $\stackrel{\circ}{\Omega}_1$ the Lyapunov function defined by $$V(S,I)=\left(S-\bar S\right)-\frac{\mu_3+\gamma}{\beta}\log\frac{-\mu_2+\mu_3+\beta S}{-\mu_2+\mu_3+\beta\bar S}+\left(I-\bar I\right)-\bar I\log\frac{I}{\bar I}$$
 It is easy to verifiy that $V$ is definite positive, that is $V(S,I)\geq0$ and $V(\bar S,\bar I)=0$ if and only if $(S,I)=(\bar S, \bar I)$. His derivative along trajectories of (\ref{2}) is given by: 
 \begin{eqnarray}
 \dot{V}(S,I) 
            &=& \dot S -\left(\mu_3+\gamma\right)\frac{\mu_3 + (\mu_2-\mu_3)I -\mu_3S-\beta SI}{-\mu_2+\mu_3+\beta S}+ \nonumber\\
          & & \;\;\;\;\;\;\; \;\;\;\;\;\;\; \;\;\;\;\;\;\; \;\;\;\;\;\;\; \;\;\;\;\;\;\;  \;\;\;\;\;\;\; \;\;\;\;\;\;\; \beta SI-\left(\mu_2 + \gamma \right)I-\bar I\left(\beta S-\left(\mu_2 + \gamma \right)\right) \nonumber\\
          &=& \dot S -\left(\mu_3+\gamma\right)\frac{(\mu_3-\mu_3S)}{-\mu_2+\mu_3+\beta S}+\left(\mu_3+\gamma\right)I+\nonumber\\
          &&\;\;\;\;\;\;\; \;\;\;\;\;\;\; \;\;\;\;\;\;\; \;\;\;\;\;\;\; \;\;\;\;\;\;\;  \;\;\;\;\;\;\; \;\;\;\;\;\;\; \beta SI-\left(\mu_2 + \gamma \right)I-\bar I\left(\beta S-\left(\mu_2 + \gamma \right)\right) \nonumber\\
          &=& \mu_3\left(1 -S\right) -\left(\mu_3+\gamma\right)\frac{(\mu_3-\mu_3S)}{-\mu_2+\mu_3+\beta S}-\bar I\left(\beta S-\left(\mu_2 + \gamma \right)\right) \nonumber\\
          &=& \mu_3\left(1 -S\right)\left[1 -\frac{\mu_3+\gamma}{-\mu_2+\mu_3+\beta S}\right]-\bar I\left(\beta S-\left(\mu_2 + \gamma \right)\right) \nonumber\\
          &=& \mu_3\left(1 -S\right)\left(\frac{-\beta\bar S+\beta S}{-\mu_2+\mu_3+\beta  S}\right)-\frac{\mu_3}{\mu_3+\gamma}\left(1-\bar S\right)\left(\beta S-\beta\bar S\right) \nonumber\\
          &=& -\mu_3\beta\left(\bar S-S\right)\left[\frac{1 -S}{-\mu_2+\mu_3+\beta S}-\frac{1-\bar S}{\mu_3+\gamma}\right] \nonumber\\
          &=& -\mu_3\beta\left(\bar S-S\right)\left[\frac{1 -S}{-\mu_2+\mu_3+\beta S}-\frac{1-\bar S}{-\mu_2+\mu_3+\beta \bar S}\right]\nonumber\\
          &=& -\frac{\beta\mu_3}{\mu_3+\gamma}\left[\frac{-\mu_2+\beta+\mu_3}{-\mu_2+\beta+\beta S}\right]\left(S-\bar S\right)^2\nonumber\\
          &\leq &0\nonumber  
 \end{eqnarray}
 Then we conclude $\dot{V}$ is semi-definite positive. Then the endemic equilibrium is stable by Lyapunov theorems. We prove the attractivity of endemic equilibrium  using Lasalle's principle.\\
 The set on which $\dot V=0$ is  given by $ E=\left\{(S,I)\in\stackrel{\circ}{\Omega}_1/\;S=\bar S\right\}.$  Then on this set, we have $\dot{S}=\mu_3 + (\mu_2-\mu_3)I -\mu_3S-\beta \bar SI=0$,  then $\displaystyle I=\frac{\mu_3-\mu_3\bar S}{\beta S-\mu_2+\mu_3}=\bar I$. Furthermore the largest invariant set contained in the set $\left\{\left(S,I\right)\in\stackrel{\circ}{\Omega}_1\;/\;\dot{V}\left(S,I\right)=0\right\}$ is reduced to the endemic equilibrium. Hence $(\bar S,\bar I)$ is attractive. Then EE is GAS on $\stackrel{\circ}{\Omega}_1$.\\
 If $\displaystyle S\leq\frac{\mu_2-\mu_3}{\beta}$, we have:
 \begin{eqnarray}
 \dot S &=& \mu_3 + (\mu_2-\mu_3)I -\mu_3S-\beta SI\nonumber\\
            &=&  \mu_3\left(1-S\right) + \left(\mu_2-\mu_3-\beta S\right)I \nonumber\\
             &>& 0\nonumber
 \end{eqnarray}
 Then $\dot S>0$. Furthermore all trajectories in $\stackrel{\circ}{\Omega}\setminus \stackrel{\circ}{\Omega}_1$ enter in $\stackrel{\circ}{\Omega}_1$. Then the set  $\stackrel{\circ}{\Omega}_1$ is absorbant. Hence the EE is GAS on  $\stackrel{\circ}{\Omega}$ .

 In the boundary    $S=0$ et $S+I=1$, the  vector field is strictly pointing inside $\Omega$. Only the $S$-axis is invariant. The endemic equilibrium is GAS on   $\Omega\backslash\left\{\left(S,I\right): I=0; 0\leq S\leq1\right\}$. This end the proof.   
 
\section{SIRS Model}
 \n
 In this section, we consider a SIRS model with different mortality. With the same notation that above, we have the following system:
 \begin{equation} \label{sirs1}
\left\{\begin{array}{lll}
\dot{S}=\Lambda-\beta\frac{S \,I}{N}-\mu_1 \,S+\nu R\\
\dot{I}=\beta\frac{S \,I}{N}-\mu_2I - \gamma \,  I\\
\dot{R}=\gamma\, I - (\mu_3+\nu) \, R 
\end{array}\right.
\end{equation}

\noindent Which reduces to 
\begin{equation} \label{sir2}
\left\{\begin{array}{lll}
\dot{S}=-\beta\frac{S\,I}{N}+\mu_2\,I+(\mu_3+\nu)\,R\\
\dot{I}=\beta\frac{S\,I}{N}-(\mu_2 + \gamma)\, I\\
\dot{R}=\gamma \,I - (\mu_3+\nu)\, R 
\end{array}\right.
\end{equation}
The system (\ref{sir2}) is exactly as system (\ref{1}) where $\mu_3$ is replaced by $\mu_3+\nu$. 
 
 \section{Conclusion}
In this contribution, we have proved the global stability of $SIR$ and $SIRS$ models with differential mortality by Lyapunov methods.
Our results encompass and improve the results of \cite{1022.34044}.

\end{document}